%% file: Main.tex
\documentclass[lettersize,journal]{IEEEtran}
\usepackage{balance}

\ifCLASSOPTIONcompsoc
  \usepackage[nocompress]{cite}
\else
  \usepackage{cite}
\fi
\ifCLASSINFOpdf
\else
\fi
\usepackage{tabularx,booktabs}
\usepackage[dvipsnames]{xcolor}
\usepackage{tikz, cite}
\usepackage{tabulary}
\usepackage{tabularx}
\usepackage{titlesec}
\usepackage{amsmath,amsfonts,color}
\usepackage[ruled,linesnumbered]{algorithm2e}
\usepackage{multirow}
\usepackage{graphicx}
\usepackage{epstopdf}
\usepackage[export]{adjustbox}
\usepackage[none]{hyphenat}
\usepackage[normalem]{ulem}
\usepackage{amsmath,amssymb}

\newif\ifdraft
\drafttrue 

\usepackage{pifont}
\newcommand{\cmark}{\ding{51}}%
\newcommand{\xmark}{\ding{55}}%

\usepackage{ragged2e}

\newcommand{\name}{{\sc M-TEC}\xspace}

\begin{document}
%
\title{Dynamic DAG-Application Scheduling for Multi-Tier Edge Computing in Heterogeneous Networks}
%
%
%
%

\author{Xiang~Li,
        Mustafa~Abdallah,
        Yuan-Yao Lou,
        Mung Chiang,
        Kwang Taik Kim,
        Saurabh Bagchi
\thanks{Manuscript received XXX; revised XXX.}}

%
%

\markboth{Journal of \LaTeX\ Class Files,~Vol.~14, No.~8, September~2024}%
{Shell \MakeLowercase{\textit{et al.}}: Bare Demo of IEEEtran.cls for Computer Society Journals}
%



\maketitle
\begin{abstract}
Edge computing is deemed a promising technique to execute latency-sensitive applications by offloading computation-intensive tasks to edge servers. Extensive research has been conducted in the field of end-device to edge server task offloading for several goals, including latency minimization, energy optimization, and resource optimization. However, few of them consider our mobile computing devices (smartphones, tablets, and laptops) to be edge devices. In this paper, we propose a novel multi-tier edge computing framework, which we refer to as \name, that aims to optimize latency, reduce the probability of failure, and optimize cost while accounting for the sporadic failure of personally owned devices and the changing network conditions. We conduct experiments with a real testbed and a real commercial CBRS 4G network, and the results indicate that \name is capable of reducing the end-to-end latency of applications by at least 8\% compared to the best baseline under a variety of network conditions, while providing reliable performance at an affordable cost.
\end{abstract}




%

\input{Introduction}
\input{ProblemStatement}
\input{Framework}
\input{Experiment}
\input{Relatedwork}
\input{Discussion}

\input{Conclusion}

\bibliographystyle{IEEEtran}
\balance
\bibliography{sample-base}

\end{document}

%% file: Introduction.tex
\section{Introduction}\label{sec:introduction}

The growing computation intensity and low latency requirement of surging applications which run on user-generated data fuels the need for moving computation closer to users~\cite{varghese2016challenges,Mung-Tao-2016}. Such requirements have led to the widespread adoption of edge computing ({EC}), which offers reduced latency by executing computations close to the data source and scalability by splitting the workload across several edge devices.

\begin{figure}[ht]
\includegraphics[width =0.9\columnwidth]{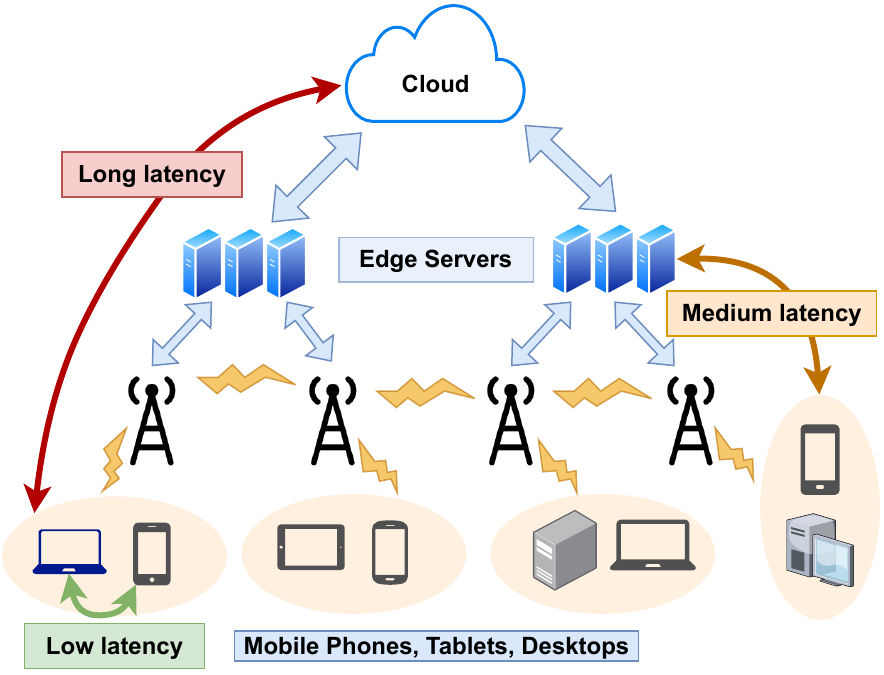}
\caption{A system overview of classical cloud computing (red), edge computing (yellow), and multi-tier edge computing with peer-to-peer offloading (green).}
\label{fig:system_overview}
\vspace{-3 mm}
\end{figure}

Figure \ref{fig:system_overview} depicts an overview of three computing paradigms: cloud computing, which is widely used in the real world, edge computing or fog computing, which is extensively discussed in the existing literature~\cite{edge_survey, edge_research, intelligent_edge, all_about_edge}, and the paradigm proposed by this work - multi-tier edge computing, which brings the computing power even closer to data source by allowing peer-to-peer offloading in close proximity.

\begin{table*}[ht]
    \setlength\tabcolsep{9pt}\renewcommand\defaultaddspace{1.1ex}
        \begin{center}
            \begin{tabular}{|c|c|c|c|c|c|c|c|}
                \hline
                \textbf{Framework} & \textbf{\shortstack{Heterogeneous\\ Edge Devices}} & \textbf{\shortstack{Low \\ Latency }} & \textbf{\shortstack{P2P \\ Offloading}} & \textbf{\shortstack{Network \\ Fluctuation }} & \textbf{\shortstack{Supporting \\ DAG }} & \textbf{\shortstack{Failure\\ Reduction}}\\
                \hline
                \cline{1-7}
                
                \name(Ours) & \cmark & \cmark &   \cmark  & \cmark & \cmark & \cmark    \\
                \hline
                IBOT~\cite{suryavansh2020bot} & \cmark & \cmark &  \xmark  & \xmark & \xmark & \cmark    \\
                \hline
                DCC~\cite{novel_work}& \cmark & \cmark &  \cmark  & \cmark & \cmark & \xmark    \\
                \hline
                LaTS~\cite{lats}& \cmark & \cmark &  \xmark  & \xmark & \cmark & \xmark    \\
                \hline
                Petrel~\cite{petrel}  & \cmark  & \cmark  &  \xmark & \xmark & \xmark & \xmark \\
                \hline
                LAVEA~\cite{lavea} & \cmark & \cmark & \xmark & \cmark & \cmark & \xmark \\
                \hline
                            
            \end{tabular}
        \end{center}
    \caption{A comparison of the available features between the prior related works and our framework (\name). }
    \label{tab:prior_work_comparison}
    \vspace{-6mm}
\end{table*}

Existing literature on EC has focused on scheduling and offloading schemes for computation-intensive services from end devices to commercially managed edge server infrastructures, under the assumption that such infrastructures will be available for extended periods and achieve reasonably low latency~\cite{mume,three_layer,cloud_mec,lavea,petrel,lats,hetmec, infocom2018, comm_letter}. Majority of the previously proposed frameworks, however, solely utilized commercially maintained edge servers. Commercially maintained edge servers have advantages in terms of dependability and performance, but they also have drawbacks. They are not yet widely available, and edge computing services are typically not free. Consequently, a framework that may leverage the omnipresent devices among us to assist with demanding tasks appears to be a viable option.

Additionally, the EC literature on device-to-device ({D2D}) offloading in terms of computing resources and data optimization is scant \cite{d2d_1,novel_work}.
Their offloading scheme only takes into account individual tasks and not the dependencies between tasks, making it less practical for real-world applications. \cite{offload1} proposed multihop offloading through edge device collaboration, in which they focus on the task dependency and formulate a fixed scheduling policy based on network flow at the beginning of the first task, which is less robust. \cite{novel_work} proposed a new three-layer framework called DCC, in which they integrate end devices into dynamic cloudlets by Kmean clustering and classify tasks into several categories to pick the optimal task allocation that aims to reduce latency and energy. However, they neglected to address the dynamic network condition of these end devices and did not account for the possibility of end device failure.

To accommodate the heterogeneous network environment and utilize the ubiquitous personal devices, we present the design of a multi-tier edge computing scheduling framework, which we refer to as \textbf{\name} 
that combines personal edge devices ({PEDs}) and commercial edge devices ({CEDs}) from different networks into a single system to leverage the benefits of both types of devices. To the best of our knowledge, \name is the first multi-tier edge computing framework that enables within-layer communication, combining end-devices, edge, and cloud from different networks into a single edge computing system to schedule complex directed acyclic graph (DAG)-based applications with low end-to-end latency, failure probability, and cost. One sample application of \name is vehicular edge computing, where tasks such as pedestrian detection, obstacle detection can be executed on board as well as other vehicles in close proximity.

\begin{figure}[t]
\includegraphics[width =\columnwidth]{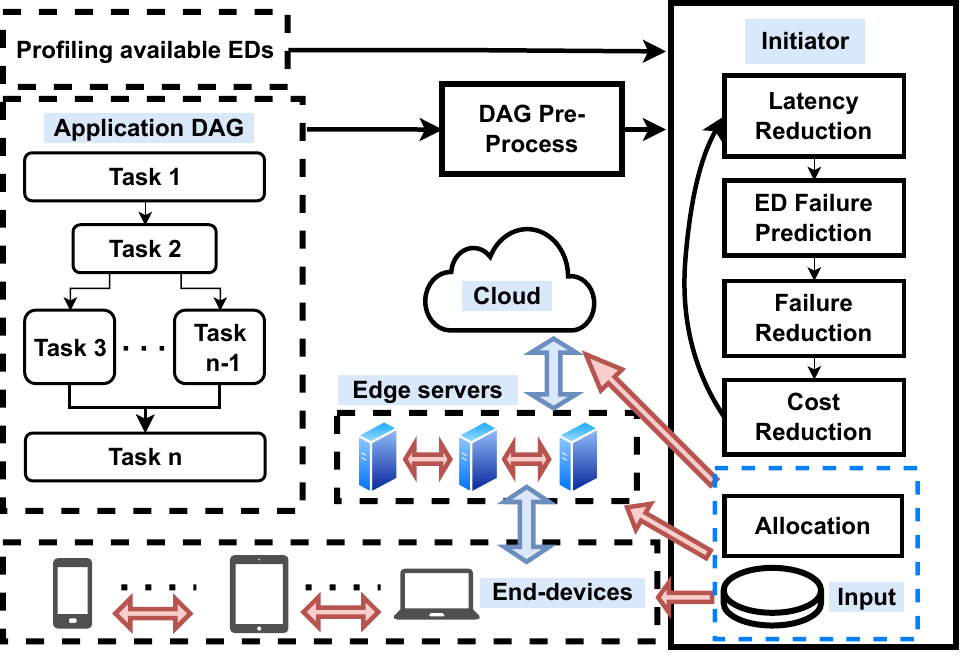}
\caption{An overview of \name. Components in dashed boxes represent profiled data or user-supplied applications in DAG form (e.g. Lightgbm as shown in the graph). The components in the solid boxes represent the orchestration framework, in which DAG preprocessing, latency reduction, failure reduction, and (\$) cost reduction are carried out.}
\label{fig:smec_overview}
\vspace{-5 mm}
\end{figure}

Figure \ref{fig:smec_overview} provides an overview of the proposed orchestration framework \name. The application instance to be offloaded needs to be represented as a directed acyclic graph (DAG) workflows, which has been addressed in previous literature~\cite{dag_rep1,dag_rep2} . The initiator preprocesses the DAG and map task dependencies (as described in detail in Section \ref{sec:dagprep}), then it employs a greedy method to lower the end-to-end latency of the application instance while taking into account the probability of failure and the dollar cost incurred. The primary contribution of this paper is the proposal of a novel three-tier dependable edge computing system that operates in a heterogeneous network environment and enables the within-layer p2p offloading to unleash the ubiquitous and powerful computing resources all around us.

In our evaluation, we compare \name to two intuitive baselines, Random Allocation and Round Robin, and four state-of-the-art solutions, LAVEA~\cite{lavea}, Petrel~\cite{petrel}, LaTS~\cite{lats}, and DCC (three-layer offloading)~\cite{novel_work}, in a commercial CBRS 4G network with real edge devices and servers, as outlined in Table \ref{tab:prior_work_comparison}. Four applications with diverse DAG structures and domains are utilized to test our framework. The results indicate that \name can lower application end-to-end latency by at least 8\% relative to the best baseline across all four testing applications across a wired or wireless connection with dynamic network traffic while reducing failure probability by more than 40\%. To the best of our knowledge, \name is the first scheduling framework for scheduling DAG workflows between peer client devices, edge servers, and cloud servers. Further, \name is deployed and tested on real devices with a real network, in contrast to simulations.

We summarize our contributions as below:

\begin{enumerate}
    \item We propose a multi-tier edge computing framework \name 
    that executes complex DAG-based user applications in a heterogeneous network that comprises both commercial edge devices and personal edge devices.
    
    \item We propose a greedy algorithm that jointly optimizes end-to-end latency, failure probability, and cost while accounting for the unpredictability of network and device failure. The fact that our algorithm is distributed is key to its deployment in a real-world environment where it is operationally difficult to rely on an always-on central scheduler. 
    
    \item We validate our framework by conducting extensive experiments in the real world network environment, which consists of Ethernet, CBRS 4G, and Wi-Fi. We demonstrate the advantages of \name in reducing average application end-to-end latency, failure probability, and cost across different networks.
\end{enumerate}

Our paper is organized as follows. In Section~\ref{sec:problem_statement}, we precisely lay out the problem. In Section~\ref{sec:design}, we present our design, first the high-level elements in our solution and then the details of each element. Section~\ref{sec:experiment} presents our extensive evaluation on the real-world testbed with real edge devices connected by a controllable CBRS 4G cellular network. We survey the relevant prior work in Section~\ref{sec:related_work} and then conclude the paper. 

%% file: ProblemStatement.tex
\section{Problem Statement} 
\label{sec:problem_statement}
The combination of commercially managed edge devices and personally owned edge devices already poses unique challenges due to the sporadic availability of personal devices. In addition, these devices are connected to separate networks, which creates an additional level of network speed, stability, and availability unpredictability. Such issues arise when attempting to combine devices from several networks to construct a dependable multi-tier edge computing platform. Such difficulties have not, to our knowledge, been addressed in the existing literature on edge computing. In this section, we will cover the key obstacles, the solutions to which will highlight the originality of our most recent work, \name.

\subsection{Preliminary and notations}
Now, we present the notations and terminology utilized by our framework \name in Table \ref{tbl:notation}. 

\begin{table}[h]
  \vspace{2mm}
  \setlength\tabcolsep{9pt}\renewcommand\defaultaddspace{1.0ex}
  \begin{tabularx}{\columnwidth}{@{}c>{\hsize=0.55\columnwidth}X>{\hsize=0.0\hsize}X @{}}
    \toprule
    \textbf{Symbols} & \textbf{Definition} \\
    \midrule
    $T = \{T_{1}, T_{2}, \hdots, T_{N}\}$ & Types of tasks in a given application  \\
    $S =\{S_{1}, \hdots, S_{N}\}, T_{i} \in S_{j}$ & Number of stages in DAG\\
    $G=(V_{i},E_{j}),\ S_{i} \in G$  & DAG representation of application\\
    \hline
    \addlinespace
    $ED =\{ED_{1}, \hdots, ED_{N}\}$ & Available edge devices (participators) in the offloading network\\
    \hline
    \addlinespace

    $L(T_{i})_{ED_{p}}$ & Execution latency of $T_{i}$ on $ED_{p}$ \\
    $L(M(T_{i}))_{ED_{p}}$ & Model download latency of $T_{i}$ \\
    $L(T_{i})_{d}$ & Data transfer latency of input for $T_{i}$ from other devices\\
    \hline
    \addlinespace
    
    $L(T_{i}),L(S_{i}), L(G)$ & End-to-end latency of task $T_{i}$, stage $i$,  application G\\
    \hline
    \addlinespace
    
    $M(T_{i})$ & Model required for $T_{i}$\\
    $H(T_{i})$ & Memory required for $T_{i}$\\
    $T(i)_{d}$ & Input data for task $T_{i}$\\
    $P(T_{i})$ & Placement of task $T_{i}$\\
    $D(T_{i})$ & Dependency of task $T_{i}$ in terms of other tasks\\
    $T_{i}(meta)$ & Meta file generated by $T_{i}$ \\
    \hline
    \addlinespace
    
    $F(T_{i})$ & Probability that $T_{i}$ fails\\
    $P_{f}(G)$ & Probability of failure of application, given by graph G\\
    $ED_{rep}(T_{i})$ & Devices that execute task $T_{i}$ replications \\
    \hline
    \addlinespace
    
    $c(ED_{p})$ & Unit time cost of edge device p\\
    $C(T_{i})$ & Cost of complete task $T_{i}$\\
    $C(G)$ & Cost of complete application instance\\
    \hline
    \addlinespace

    $T_{rep}$ & Tracker for the number of replications\\
    $\phi$ & Probability of failure threshold\\
    $\kappa$ & Threshold of the Replication degree\\
    $\alpha, \beta, \gamma$ & Hyper-parameter for tuning the weight on latency, probability of failure, and cost \\
    $\eta$ & Hyper-parameter for transmission speed error over provision \\
    \hline
    \addlinespace

    $P(G)$ & Placement of each task in graph G\\
    \hline
    \addlinespace
    $Ed_{info}$ &  Total and free space on each ED\\ 
    $M_{info}$ & Available models on each edge device\\
    $Task_{info}$ & Data structure to track executing tasks and types on each ED  \\
    
    $WeightS$ & Weight score of joint optimization \\
    $WeightSnew$ & Weighted score after PF reduction\\
    \hline
    \addlinespace
  \end{tabularx}
  \caption{Summary of the notations and their respective definitions used in \name. To improve readability, we arrange symbols that are linked to a task (or function) together.}
  \label{tbl:notation}
  \vspace{-5 mm}
\end{table}

\noindent\textbf{Peer-to-peer offloading:}
The majority of existing literature on task offloading/scheduling focuses on client-edge or client-cloud offloading, assuming the dependability and performance of commercially managed edge servers. However, such edge servers have not been widely deployed, and they are relatively expensive. Therefore, a multi-tier edge computing framework that permits peer-to-peer offloading in the client tier to release the power of idle computing resources, such as desktops, laptops, tablets, etc., appears promising.

\noindent\textbf{Heterogeneity in the devices and communication network among devices:}
Different devices have different processing power, memory, etc and they are connected to various networks. Mobile phones and tablets can interact with one another over Wi-Fi or cellular network, whilst desktop computers and servers communicate with one another via LAN. As the network connection of privately owned devices might vary, heterogeneity in network connection adds an additional layer of complexity to addressing dynamic network conditions.

\noindent\textbf{Dynamic network conditions:} Dynamic network conditions, which relate to the ever-changing data transmission speed and latency among devices and between devices and servers, make it difficult to establish a stable allocation scheme that ensures consistent performance. To orchestrate the execution strategy optimally, a dynamic orchestration framework that monitors network traffic between every two edge devices within the orchestration network is required. Such an approach requires the initiator to be aware of the global network condition and, if not managed effectively, can incur excessive latency overhead.

\noindent\textbf{Application's DAG:} An application could be made up of a few different tasks, and there might be data dependencies or order of execution requirements between those tasks. This dependency adds another layer of complexity, as some tasks can be conducted in parallel while others must be executed in a specific order. The remainder of this paper uses DAG to represent an application, where a node of the DAG represents a task in an application and the edge connecting the nodes reflects their dependency. The smallest unit used for scheduling in this paper is a task (node). Static DAG partition method is used throughout the paper.

%% file: Framework.tex
\section{Design of \name}
\label{sec:design}

In order to address the problems outlined in the preceding section, we propose a multi-tier edge computing framework that we refer to as \name. \name is a dynamic  decentralized scheduling framework for complex DAG-based applications that aims to jointly optimize the end-to-end latency, probability of failure, and cost based on the application's requirements while taking into account the uncertain nature of different networks and the intermittent availability of edge devices. The remainder of this section discusses the framework's design components and our proposed algorithm for scheduling.

\subsection{Design components}
Each device in the network has two sets of functional programs - the initiator program and the participator program. Depending on the user's intention of joining the network, it will select the appropriate program to execute. If the user joins the network as an initiator, it indicates that the user is intended to offload some tasks. Therefore, the initiator is responsible for {DAG-Preprocessing} and collecting the profiling information from each device in the network for optimization purpose. As each initiator only process the offloading requests originated from itself, it will not be the bottleneck of the system.

If the user joins the network as a participator, it indicates that the user is willing to contribute its idle resources. It will inform the initiator whether the task to be offloaded has been profiled on itself. If it is not, it should be profiled before it can join the offloading network.

\begin{figure}[h]
\includegraphics[width =\columnwidth]{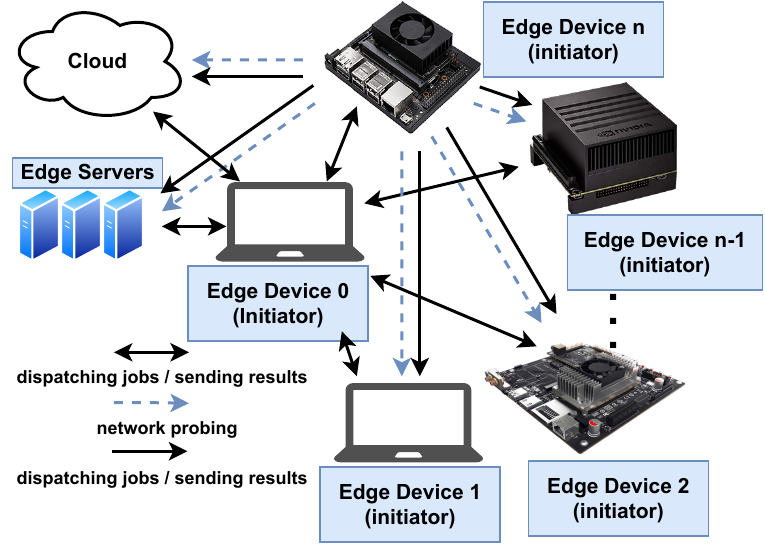}
\caption{A brief overview of the offloading network. Note that there isn't a fixed initiator in the network and the role of initiator and participator can be switched. There can be multiple initiators in the offloading network.}
\label{fig:two_role}
\vspace{-5mm}
\end{figure}

\subsubsection{\textbf{DAG-preprocessing}} \label{sec:dagprep}~\\
When the initiator decides to offload an application instance, it needs to transform the application's DAG and divides the execution into stages. The benefit of breaking the DAG into stages is that the task dependencies are integrated inside the stages, and all tasks within the same stage can be done concurrently. This DAG transformation is carried out using a modified version of Breadth-First Search, in which the stage of a node is the length of the longest path from the start node.

\subsubsection{\textbf{Initiator}} \label{sec:orch}~\\
The initiator is the origin of the offloading process. When a device join the network as an initiator, it quickly discovers the available participators who already in the network around itself and gathering the profiling information on those devices against the task that is going to be offloaded. If such information on a participator is not available, the initiator can request profiling information from a similar device in the network. If such a similar device does not exists, the participator will be removed from the current offloading process and put on profiling. It will rejoin the offloading network once the profiling is done. We use the linear, independent, and additive property~\cite{ours} of collocated task for latency estimation, which is described in Sec~\ref{subsec: interference}.

\noindent \textbf{End-to-end latency:} With the profiling information of each participator available, the initiator estimates the end-to-end latency of each task in the DAG-based application on each participator $p$ in the offloading network, which is defined as $ L(T_{i})_{ED_{p}}$. The estimated end-to-end latency of executing task $T_{i}$ on each participator is rated from low to high with their assigned edge device, and this procedure is described as follows:
\begin{align} \label{eqn: latency}
& \underset{p}{\arg \min }~ L(T_i) \nonumber \\  
& \text{where} \ L(T_i) = L(T_{i})_{ED_p}+L(M(T_{i}))_{ED_{p}}+L(T_{i})_{d} \\
& \text{s.t.} \ H(T_{i}) \leq H(ED_{p}) \nonumber, \ ED_{p}\in ED
\end{align}
Here, the end-to-end latency of $T_{i}$ consists of three components: 
task execution latency $L(T_{i})_{ED_p}$, inferred from profiling data and the number of distinct types of tasks co-located on the assigned device and tracked by the structure $\text{Task}_{\text{info}}$; 
model downloading latency $L(M(T_{i}))_{ED_{p}}$, which is determined based on model size and network download speed, if the task requires the execution of a model; 
and input data transmission latency $L(T_{i})_{d}$, which is determined depending on the input file size and the transmission speed between source and destination devices. 
$H(T_{i})$ is the memory required for $T_{i}$'s execution, including memory to store data and model, whereas $H(ED_{p})$ is the memory available on $ED_{p}$, which is tracked by $ED_{\text{info}}$. The aforementioned procedure is illustrated in Algorithm 1. 
A priority queue is used to store the allocation of task $T_{i}$, with devices in ascending order of lowest latency to highest latency.

Since the network conditions change dynamically, the transmission speed between devices also varies accordingly. This fluctuation in transmission speed is accounted for in our framework, as described in Section \ref{sec:client}.

Let us now define $L(S_{i}) = \max_{T_i \in S_i} L(T_{i})$ as the latency for stage $i$, where $L(T_{i})$ is given by \eqref{eqn: latency}. As a result, the total end-to-end latency of the application is equal to the sum of the latencies of the tasks in each stage that take the longest:
\begin{equation*}
    L(G) = \Sigma_{i=1}^{i=S}L(S_{i}). 
\end{equation*}

\begin{algorithm}
    \SetAlgoLined
    \caption{Min\_Latency\_Scheduling}
    \SetKwInOut{KwIn}{Input}
    \SetKwInOut{KwOut}{Output}
    \SetKwInOut{KwInit}{Initialization}
    \SetKwInOut{KwRe}{return}
    \KwIn{$T_{i}$}
    \KwOut{$L(T_{i})-Queue$}
    \For{$ED_{p} \in ED$ }{
        $L(T_{i})_{ED_{p}} =
        GetEstimatedTime(T_{i},ED_{p})$\\
        $L(M(T_{i}))_{ED_{p}} = 0$\\
        \If{$M(T_{i})\ not\ on\ {ED_{p}}$}{
           $L(M(T_{i}))_{ED_{p}}=
            GetModelDownloadTime(size(M(T_{i})),B)$\\
        }
            \If{$T(i)_{d}\ not\ on\ \ ED_{p}$}{
        $L(T_{i})_{d} = GetDataTransTime(T(i))$\\
        }
        $L(T_{i})=L(T_{i})_{ED_{p}}+L(M(T_{i}))_{ED_{p}} + L(T_{i})_{d}$\\ 
            $L(T_{i})\_Queue.enqueue([ED_{p},L(T_{i})])$
        }  
    \label{algo:latency_reduction}

\end{algorithm}
\vspace{-2mm}

\noindent \textbf{Probability of failure:} Now, with the end-to-end latency of the task $T_{i}$ executing on each available participator, the initiator attempts to reduce failure probability and cost. Due to the unmanaged nature of personal devices, their availability may fluctuate. Therefore, it is vital to incorporate redundancy to replicate jobs given to participators with a high failure probability. To estimate the failure probability of each participator, we construct a failure prediction model for each participator, which is expressed as the exponential function~\cite{mobility,ours} given by 
\begin{align*}
    P(ED_{i}) = 1 - e^{-\lambda t}, 
\end{align*}
where the failure rate $\lambda$ is estimated from the  participator's history data. Since tasks fail as soon as those devices they are assigned to fail, therefore, the probability of failure for a task $T_{i}$ equals the probability of failure of the participator that executes the task $T_{i}$ during its execution period, which is denoted by $F(T_{i})$. 

Now, the initiator estimates the failure probability of the most optimal allocation for end-to-end latency for task $T_{i}$ determined in the previous phase. If $F(T_{i})$ is greater than a certain threshold $\phi$ and the number of replication for task $T_{i}$ is less than the maximum number of replications allowed $\kappa$, a weighted score, $WeightS$, is computed for the current allocation for $T_{i}$, with user-defined weights $\alpha,\beta,\gamma$ assigned to end-to-end latency, probability of failure, and cost.
The framework then attempts to replicate tasks on the second-most optimal option from the end-to-end latency queue in an effort to lower the failure probability. A new weighted score is derived from the new latency, failure probability, and cost. If the new weighted score, $WeightSnew$, is less than the original weighted score, $WeightS$, task $T_{i}$ is replicated on the second most optimal allocation. The process is continued until $F(T_{i})$ is less than the failure threshold $\phi$, the total number of replications exceeds the maximum allowed replications or $WeightS$ is no larger than $WeightSnew$. All participators that execute replications of task $T_{i}$ are denoted by the set $ED_{rep}(T_{i})$. The above-described procedure is depicted in Algorithm 2. 

The entire application instance is deemed successfully completed when every task \( T_i \) in the instance is successfully executed, denoted as \( T_i^s \). The failure probability for the application instance is given by:
\[
P_f(G) = 1 - \Pr(T_1^s, T_2^s, \ldots, T_N^s).
\]
Given that the tasks in the application instance are structured as a DAG, the events of successful task completions are conditional on the completions of prior tasks. To compute the probability of failure for a given application instance, consider the following example where \( A \rightarrow B \) implies that task \( B \) depends on task \( A \). For a DAG application \( G \) with six tasks where the dependencies are \( T_0 \rightarrow T_1 \), \( T_1 \rightarrow T_3 \), \( T_1 \rightarrow T_4 \), \( T_0 \rightarrow T_2 \), and \( T_2 \rightarrow T_5 \), the probability of failure is calculated as:
\[
\begin{aligned}
    P_f(G) &= 1 - \Pr(T_0^s) \Pr(T_1^s \mid T_0^s) \Pr(T_3^s \mid T_1^s, T_0^s) \\
    & \cdot \Pr(T_4^s \mid T_1^s, T_0^s) \Pr(T_2^s \mid T_0^s) \Pr(T_5^s \mid T_2^s, T_0^s).
\end{aligned}
\]

\begin{algorithm}
    \SetAlgoLined
    \caption{PF\_Cost\_Reduction}
    \SetKwInOut{KwIn}{Input}
    \SetKwInOut{KwOut}{Output}
    \SetKwInOut{KwInit}{Initialization}
    \SetKwInOut{KwRe}{return}
    \KwIn{$L(T_{i})\_Queue$}
    \KwOut{$P(T_{i})$}
    $F(T_{i}) = GetPf(T_{i},ED_{p},L(T_{i})_{ED_{p}},D(T_{i}))$\\
    $WeightS=\alpha L(T_{i})+\beta F(T_{i}) + \tau C(T_{i})$\\
    \While{$F(T_{i}) \geq \phi$ and $T_{rep}<\gamma$}{
        $ED_{p},L(T_{i} = L(T_{i})\_Queue.dequeue$\\
        $F(T_{i}) = GetPf(T_{i},ED_{p},L(T_{i}))$\\
        $C(T_{i}) = L(T_{i})*c(ED_{p}) $\\
        $WeightSnew=\alpha L(T_{i})+\beta F(T_{i}) +\gamma C(T_{i})$\\
       \If{$WeightSnew \leq WeightS$}{
            $ED_{rep}(T_{i}).add(ED_{p})$\\
            $WeightS = WeightSnew$\\
            $T_{rep}++$\\
            }{
                 }
            }
\label{algo:failure_cost_reduction}
\end{algorithm}

\noindent \textbf{Cost:} With end-to-end latency and failure probability taken into account, the framework's final optimization objective is cost. If replication is not required, the cost for the task $T_{i}$ executing on participator $p$ is determined as follows:
\begin{align*}
    C(T_{i}) = L(T_{i})_{ED_p}*c(ED_{p})
\end{align*}
where $c(ED_p)$ is the unit time cost for using  participator p.
If replications exist, the overall cost for task $T_{i}$ is determined as the sum of all costs incurred by each participator that executes task $T_{i}$, which can be represented as follows:
\begin{align*}
 &   C(T_{i}) = \sum_{n=1}^\text{\#\ of\ replications} c(ED_{p_{n}})*L(T_{i})_{ED_{p_{n}}} \\
 &  \text{s.t.}\ ED_{p_{n}} \in ED_{rep}(T_{i}).   
\end{align*}
The total cost of the entire application is given by 
\begin{align*}
    C(G) = \sum_{i=1}^{N}C(T_{i}).
\end{align*}

The final optimization problem can be expressed as:
\begin{align}
& \min  \alpha L(G)+ \beta P_{f}(G) + \gamma C(G) \nonumber \\
&  s.t.\ \alpha + \beta + \gamma = 1
\label{eqn:joint_optimization}
\end{align}
In this setup, the parameters $\alpha$, $\beta$, and $\gamma$ are user-defined weights that control the trade-offs among end-to-end latency, failure probability, and cost. This approach allows for fine-tuning based on the specific application requirements. Such joint optimization problems are commonly addressed in the literature by assigning different weights to linearly combined metrics, as seen in references like~\cite{deep_decision, eg_cost, lt_quality, ser_orien}.

\subsubsection{\textbf{Participators}} \label{sec:client}~\\
In the following, we detail the role of a participator within the offloading network. When a user joins as a participator, they consent to allocate their idle computational resources to the initiators within the network. A participator is restricted to involvement in a single offloading network concurrently. The key function of a participator is to execute tasks that are dispatched by the initiator.
Moreover, each participator must also monitor the communication speed between itself and other participators in the network (as depicted by the blue dashed arrows in Figure \ref{fig:two_role}). The transmission rate between edge participators $i$ and $j$ is measured as follows:
\begin{align*}
    S_{(i,j)}  = \frac{2*\text{size}(x)}{\text{rtt}_{(i,j)} * \eta},
\end{align*}
where $\text{size}(x)$ represents the testing packet's size, 
$\text{rtt}_{(i,j)}$ represents the round-trip time required for the testing packet to be transferred from device $i$ to device $j$ and back to device $i$, and 
$\eta$ is the transmission speed error over-provision hyper-parameter. 
The reason for including such a parameter is that the current observed network speed is used to orchestrate future tasks, and by the time offloading occurs, there should already be tasks running on the device, which can cause packet queuing and packet processing delays. This parameter provides a lower bound for the transmission speed and inhibits task offloading that could result in lengthy transmission latency. One may believe that such transmission error over-provisioning could lead to the aggregation of tasks on particular devices. We discuss the specifics of this issue in Section \ref{sec:eta_eva}. The $network\_probe$ function is shown in Algorithm 3.


\begin{algorithm}
    \SetAlgoLined
    \caption{network\_probe}
    \SetKwInOut{KwIn}{Input}
    \SetKwInOut{KwOut}{Output}
    \SetKwInOut{KwInit}{Initialization}
    \SetKwInOut{KwRe}{return}
    \KwIn{ED, $\eta$}
    \KwRe{$S_{i}$}
    \For{$ED_{p} \in ED$}{
        $\text{rtt}_{(i,p)}=send\_test\_packet(x)$\\
        $S_{(i,p)}$ $=transmission\_speed\_calculation$$(\text{rtt}\_{(i,p)},\eta)$
    }
\label{algo:network_probe}
\end{algorithm}

Additionally, participants collect input and output data size for their assigned tasks, which is sent to the initiator to build and continuously update a regression model for predicting intermediate file sizes in future orchestration~\cite{sonic}.

\subsection{Scheduling algorithm}
We have now reviewed the framework's specifics. Let's go over the deployment overview. Each device in the offloading network can potentially run two algorithms, which we refer to as the initiator and the participator. The initiator originates offloading requests, pre-processes the DAG, and optimizes latency, probability of failure, and cost before dispatching tasks to participators in accordance with the optimization allocation scheme. Algorithm 4 illustrates the initiator algorithm.

\begin{algorithm}
    \SetAlgoLined
    \caption{Initiator}
    \SetKwInOut{KwIn}{Input}
    \SetKwInOut{KwOut}{Output}
    \SetKwInOut{KwInit}{Initialization}
    \SetKwInOut{KwRe}{return}
    \KwIn{DAG-based application G , $\alpha, \beta, \tau, \phi$}
    $S=Pre\_process(G)$\\
    \For{$S_{i} \in S$}{
        \For{$T_{i} \in S_{i}$}{
            $L(T_{i})\_Queue=Min\_Latency\_Orchestration(T_{i},S)$\\
            $P(T_{i})=PF\_Cost\_Reduction(L(T_{i})\_Queue)$
        }
    }
    $Dispatch(P(G),ED)$\\
    \# following functions running in separate threads \\
    $S=Receive\_Network\_Test\_Result()$\\
    $Receive\_InOut\_Size\_Result()$\\
    $Update\_Meta\_Size\_Regression()$\\
\label{algo:orchestrator}
\end{algorithm}

The participator receives tasks and an allocation scheme from the initiator, as well as task inputs from other clients or the initiator. As soon as all necessary input files are ready, clients begin completing their given tasks. 
Algorithm 5 depicts the general function of the participator.

\begin{algorithm}
    \SetAlgoLined
    \caption{Participators}
    \SetKwInOut{KwIn}{Input}
    \SetKwInOut{KwOut}{Output}
    \SetKwInOut{KwInit}{Initialization}
    \SetKwInOut{KwRe}{return}
    \KwIn{P(G), ED, $\eta$}
    $T_{i}=Receive\_job()$\\
    $D(T_{i})=Receive\_inputs()$\\
    $T_{i}(meta) = Execute\_job(T_{i},D(T_{i}))$\\
    \# following function running in separate threads\\
    $S = network\_probe(ED)$\\
    $Send\_network\_result(S)$\\
    $Send\_InOut\_size(T_{i})$\\
\label{algo:client}
\end{algorithm}

\textbf{Everyone can be initiator:} Note that our framework does not rely on a centralized server to coordinate the scheduling process. Every initiator who intended to offload tasks will have their own smaller scaled offloading network. 
In the event that multiple initiators are present, however, it must be assured that those smaller scaled offloading networks do not overlap. For instance, if we have two initiators with their respective offloading netwotk with participators $offNet1=\{ED_{a1}, ED_{a2}, \hdots, ED_{an}\}$ and $offNet2=\{ED_{b1}, ED_{b2}, \hdots, ED_{bn}\}$, then the following condition must be true:
\begin{align*}
    offNet1 \cap offNet2 = \emptyset.
\end{align*}
The participators can be regrouped into different offloading networks when one initiator completed its tasks and a new initiator joins the network.

\subsection{Interference based service time prediction}
\label{subsec: interference}

Previous literature demonstrated that co-located tasks on the same device interfere with one another and that this interference increases the total execution time of the tasks~\cite{suryavansh2020bot, ours}. Specifically, \cite{ours} demonstrated that there is a correlation between the execution latency of a task and the number of different types of other tasks already running on the device, and they validated this correlation on multiple platforms with various operating systems and architectures. We take a similar approach in our work, profiling our testing applications against all candidate devices in the testbed and characterizing the relationship between execution latency and the number of different types of tasks executing on the devices. We present a linear regression model to capture this relationship and utilize it to predict the execution time of incoming tasks. Note that a device will not be considered as the candidate of the orchestration network until its profiling is finished. Once the device is profiled, its profiling data will be shared with other devices in the network to avoid redundant profiling.

\begin{figure*}[ht]
\includegraphics[width=1.65\columnwidth]{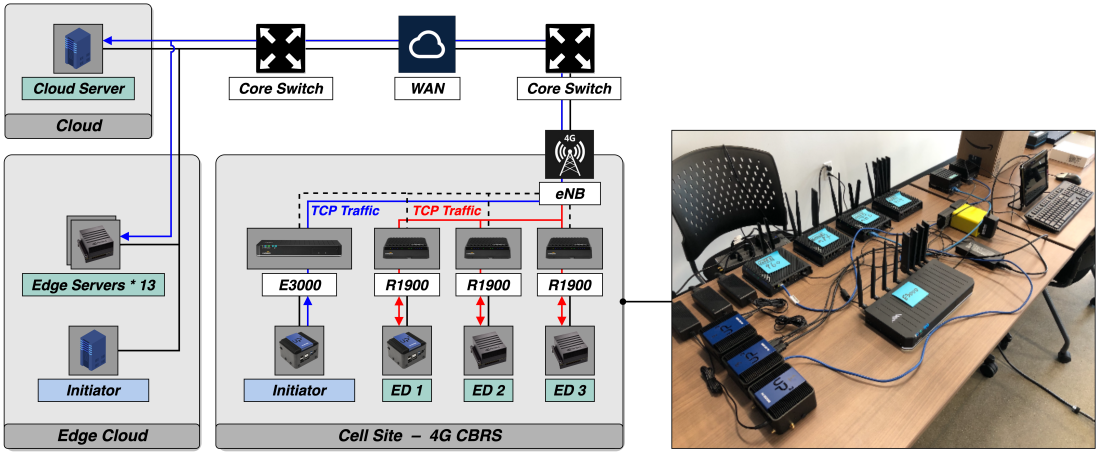}
\caption{The overview of testbed setup with the TCP background traffic flow and the physical view of testbed at the cell site..}
\label{fig:testbed}
\vspace{-4 mm}
\end{figure*}

\subsection{Scalability, privacy, and fairness}
A main disadvantage of the existing centralized scheduling approach lies at its scalability. 
Due to the large amount of devices existing in the network, it is impossible to check all of them and schedule a task in an efficient manner. \name works around this disadvantage by letting the initiators divide the large network into smaller scaled self-governed offloading networks based on the proximity, thus significantly reducing the scheduling overhead. The proximity is measured as the round trip time from the initiator to the ED. The ideas of dividing the entire orchestration network into subnets are used by \cite{cluster_head1, cluster_head2} as well. However, instead of using round trip time to measure proximity, they use devices with localization capability as gateways to manage other devices on the same local area network. Such methods did not take network traffic into consideration, which may result extra orchestration overhead.

Privacy is another concern when dealing with personally owned devices. There exists a large body of literature focus on data privacy, user location privacy and storage privacy in a distributed system~\cite{privacy1,privacy2,privacy3,privacy4}, which is beyond the scope of this paper. Such privacy preserving tasks offloading techniques have been proven feasible in practice such as in the field of federated learning. 

There exists rich literature that focuses on fairness scheduling on edge and cloud computing~\cite{fairness1,fairness2,fairness3,fairness4}. That is not a focus of this paper. However, fairness can be easily enforced by each initiator in the network during the orchestration stage.

%% file: Experiment.tex

\section{Experiment} 
\label{sec:experiment} 

To assess our framework in diverse real-world scenarios, we construct a testbed comprising a CBRS 4G network and enterprise wireless routers (Cradlepoint E3000 and R1900) connected to heterogeneous edge devices. 
Our testbed, unlike simulation-based testbeds, does not contain any assumptions or fixed network settings (e.g., end-to-end latency, data rate, queueing delay, etc.). 
Instead, our testbed can connect to a CBRS 4G network to experience and assess the actual impact of wireless channel dynamics.



In contrast to the private CBRS 4G network, we were unable to conduct experiments on commercial 5G networks because of the inability to configure static and public IP addresses for Cradlepoint gateways (which provide cellular connections to edge devices) and inbound connections for specific external IP addresses of edge devices/servers located outside of premise without the support of an operator. This prevented us from being able to use commercial 5G networks for our research.
Nevertheless, our results conducted over CBRS 4G networks will be comparable to those conducted over 5G networks with sub-6GHz frequency bands, as almost all operators currently adhere to the NSA (non-standalone) architecture in which the 5G RAN (radio access network) is connected to a 4G LTE core network~\cite{3gpp-rel15}, and channel characteristics between the CBRS frequency band at 3.5GHz and sub-6GHz frequency bands are comparable.

\begin{table*}[ht]
    \vspace{2mm}\setlength\tabcolsep{9pt}\renewcommand\defaultaddspace{1.1ex}
        \begin{center}
            \begin{tabular}{|c|c|c|c|c|c|c|}
                \hline
                \textbf{Location}           & \textbf{Device}               & \textbf{Name}     & \textbf{Amount} & \textbf{CPU} & \textbf{RAM}                     \\
                \hline
                
                Cloud                       & Cloud Server                  & Desktop           & 1 & \multirow{2}{*}{16 core (x86)}    & \multirow{2}{*}{32GB}     \\
                \cline{1-4}
                
                \multirow{4}{*}{Edge Cloud} & Edge Server                  & Desktop           & 1 &                                   &                           \\
                \cline{2-6}
                            
                                            & \multirow{3}{*}{Edge Servers} & Jetson AGX Xavier & 2 & 8 core (ARM)                      & 16 GB                     \\
                \cline{3-6}
                
                                            &                               & Jetson Xavier NX  & 7 & 6 core (ARM)                      & 8 GB                      \\
                \cline{3-6}

                                            &                               & Jetson TX2        & 4 & 4 core (ARM)                      & 8 GB                      \\
                \hline
                
                \multirow{4}{*}{Cell Site}  & Initiator (ED)  & \multirow{2}{*}{Up Squared AI Edge} 
                                & \multirow{2}{*}{2} & \multirow{2}{*}{4 core (x86)} & \multirow{2}{*}{8 GB}     \\
                \cline{2-2}
                
                                            & ED 1                          &                   &   &                                   &                           \\
                \cline{2-6}
                
                                            & ED 2  & \multirow{2}{*}{Jetson AGX Xavier} & \multirow{2}{*}{2} & \multirow{2}{*}{8 core (ARM)} & \multirow{2}{*}{16 GB}   \\
                \cline{2-2}
                
                & ED 3  &   &      &   &   \\
                \hline
            \end{tabular}
        \end{center}
    \caption{The hardware and physical information of the testbed.}
    \label{tab:testbed}
    \vspace{-8.5mm}
\end{table*}

\subsection{Experiment scenarios}
To evaluate our system, we design and conduct three distinct types of real-world experiments. 
\begin{itemize}
    \item Offloading experiments for two-tier edge cloud and cloud scenarios that connect through wired internet, aim to quantify the relationship between end-to-end latency and the number of edge loud devices involved in the offloading.
    \item Offloading experiments for three-tier cloud, edge cloud, and cell site that connects over wired, wireless link, aim to show the feasibility of using cell site devices for computation offloading.
    \item Offloading experiments with various network dynamics to examine the performance of task offloading with realistic background TCP traffic.
\end{itemize}


Figure \ref{fig:testbed} depicts an overview of the testbed setup and experiment system architecture. The testbed is divided into three tiers. The cloud tier consists of one powerful PC, the edge cloud tier is a combination of edge devices and edge servers, and the cell site is a combination of user devices.


\subsubsection{\textbf{Two-tier edge computing at edge and cloud}} \label{two_tier_exp} ~\\
First, we consider an Ethernet-based wired interconnection in an edge cloud that adheres to the traditional cloud model in order to offload tasks both at the edge and in the cloud. In addition, given the fluctuating computational resources, there are four experiment iterations. We randomly select a different number of edge servers in each iteration (i.e., 5, 8, 11 and 14). The heterogeneity of cloud and edge servers given in Table \ref{tab:testbed} reveals four distinct hardware specifications, including desktop, Jetson AGX Xavier, Jetson Xavier NX, and Jetson TX2. All communications between endpoints within the edge cloud (e.g., TCP connection setup, data transmission, etc.) are conducted via wired lines.

\subsubsection{\textbf{Three-tier edge computing in CBRS 4G network}} \label{three_tier_exp} ~\\
Instead of depending just on the power of the edge cloud and cloud server, we construct a three-tier edge computing system by incorporating a CBRS 4G network-operated cell site. 
The heterogeneity of the cell cite is increased by the presence of four devices with two distinct types (i.e., Up Squared AI Edge and Jetson AGX Xavier).
In contrast, from a network perspective, end-device communication now utilizes both wireless and wired connections. 
This is an example of how mobile edge computing typically facilitates compute offloading for end users. 
In addition, we decrease the number of selected edge servers in the edge cloud to a maximum of five in order to observe the performance of each offloading scheme on the wireless connections. 
This increases the likelihood that tasks will be dispersed among edge devices at the cell site, hence enhancing the potential for additional data transmission over wireless networks.
    
\subsubsection{\textbf{Wireless channel dynamics and TCP traffic flow}} \label{traffic_exp} ~\\
Extending the second experiment, we evaluate the performance of several schemes and underlying theories in a real-world network environment with variable network conditions. By having edge devices continuously transmit and receive TCP packets from one another, we introduce TCP background traffic into the testbed. Under this configuration, eNodeB (eNB) uplink and downlink channels would be affected most effectively. In addition, we cause the initiator to continue sending TCP packets to edge servers to simulate different types of background traffic. The TCP traffic flows are depicted in Figure \ref{fig:testbed}.

In a nutshell, the cloud and mobile edge computing paradigms, as well as a realistic edge computing scenario with customizable network traffic control, are respectively represented by the three experiments. In the following section, we provide the essential setup for conducting these real-world experiments on our testbed.

\subsection{\textbf{Testbed setup}} \label{sec:testbed}
At the edge cloud, we deploy edge servers with Ethernet connections in accordance with the standard data center architecture. The cloud server is not co-located with the edge servers, even if they are installed in the same region. Under the support of network administrators, both cloud servers and edge servers can simply acquire a static and public IP address. In contrast, at the cell site, we individually connect four end devices to Cradlepoint enterprise wireless routers. Wireless routers have SIMs for CBRS 4G network access. The edge cloud, wireless routers, and a CBRS 4G network are then configured to provide connected devices with static and public IP addresses. To accomplish computation offloading, we enable peer-to-peer (p2p) connections between edge devices. Furthermore, we construct a three-tier mobile edge computing testbed that operates on a CBRS 4G network. Note that the initiator in our testbed can be any edge device, including edge servers. This is a plausible scenario in which any of us might require computation offloading.

Next, with the configuration described above, we populate the testbed with TCP traffic traveling from end to end. As explained in Section \ref{traffic_exp}, in order to observe the impact of wireless channel dynamics on our framework, edge devices at the cell site continually send TCP packets to each other, and the initiator continuously transmits TCP packets to edge servers in the edge cloud. To bring network traffic into our testbed, we install the Linux tool \textit{iperf3} on each end device, which subsequently hosts an iperf3 session to receive TCP packets from corresponding clients. The TCP session's congestion control algorithm is \textit{cubic}, and the TCP traffic flows are depicted in Figure \ref{fig:testbed}.

The construction of a testbed on a CBRS 4G network to accommodate three types of realistic experiments has been discussed. Additionally, we noted the paradigm or scenario that each experiment illustrates. In the following section, we evaluate our framework and other compute offloading schemes on our testbed using the aforementioned three experiments. We evaluate performance from multiple perspectives, including end-to-end latency, failure probability, and financial cost. 

\subsection{\textbf{Performance metrics}}
\textbf{End-to-end latency:} We define the end-to-end latency of an application instance as the time between the dispatch of the initial task and the receipt of the final result. In our evaluation, application instances may arrive in a clustered fashion, which may result in tasks accumulating on edge devices and longer end-to-end latency for some instances. As a result, in all of our evaluations, we use the average end-to-end latency of application instances.

\noindent \textbf{Probability of failure (PF):} The probability of failure for an application instance is the probability that it does not complete all of its tasks successfully and fails to return the result to the initiator. Due to sporadic availability of edge devices or excessively extended execution times, tasks may fail (e.g., a person leaves the room with his laptop during the middle of the task execution).

\noindent \textbf{Dollar cost (\$):} Using commercially managed edge devices comes at a cost. Throughout our evaluation, we assign dollar costs to each device using a similar pricing structure as Amazon EC2 instances~\cite{ec2_price}.

\subsection{\textbf{Testing applications}}
\begin{figure}[t]
\includegraphics[width=0.95\columnwidth]{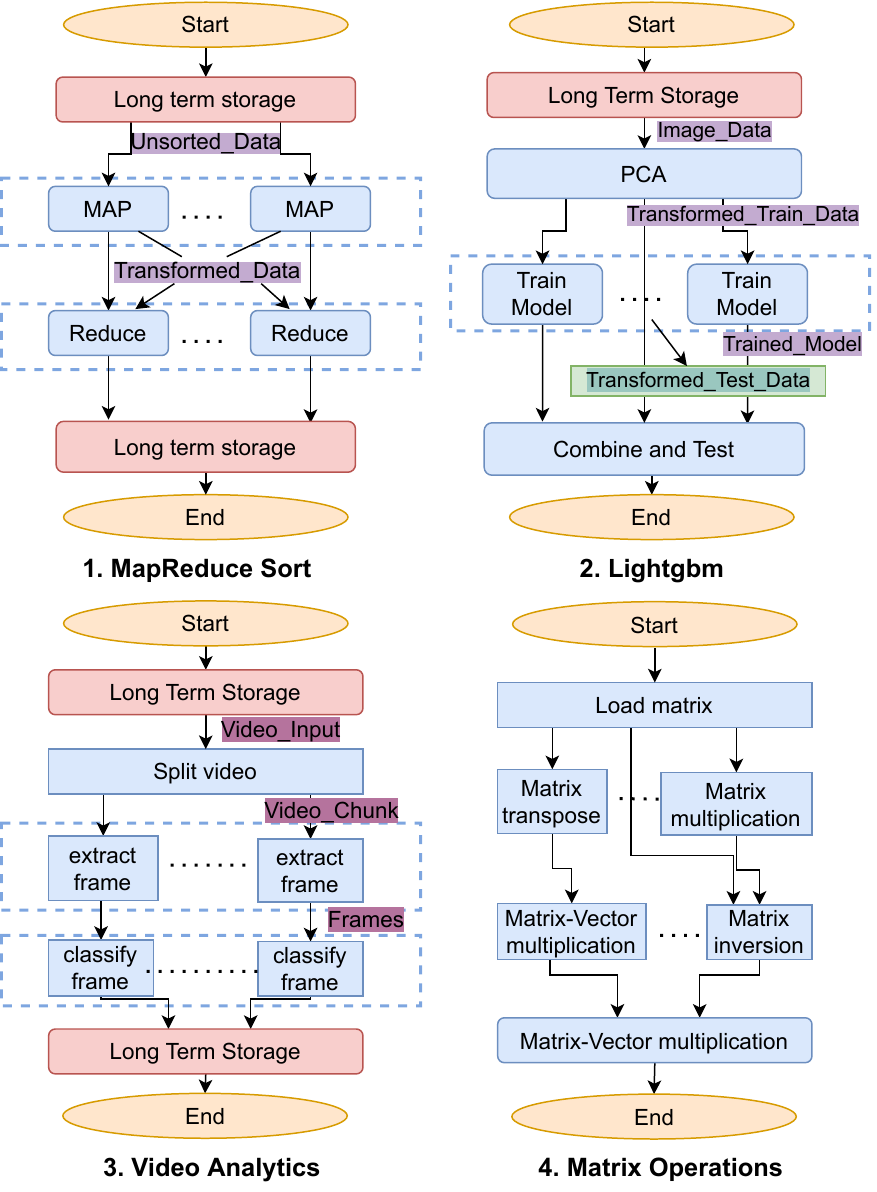}
\caption{DAG-based applications used in the experiments.}
\label{fig:test_app}
\vspace{-5 mm}
\end{figure}

We evaluated the performance of \name using four applications from diverse domains, including data science (mapreduce sort), machine learning (LightGBM, video analytics), and mathematics (matrix computation).
The DAG structure of these four applications is depicted in Figure \ref{fig:test_app}. \textbf{MapReduced Sort}: Parallel mappers fetch input job and generate intermediate files, which are consumed by reducer to produce the final result. \textbf{Lightgbm}: Train several decision trees in parallel and combine them to form a random forest predictor. \textbf{Video analytics}: split videos into batches and process them in parallel to generate analytical results. \textbf{Matrix operation:} Heavy matrix computation. Such four applications span a variety of dependency levels among tasks that are used for testing the generality of \name.

\subsection{\textbf{Latency}}

\subsubsection{\textbf{End-to-end latency comparison}}

\begin{figure}[ht]
\includegraphics[width=0.95\columnwidth]{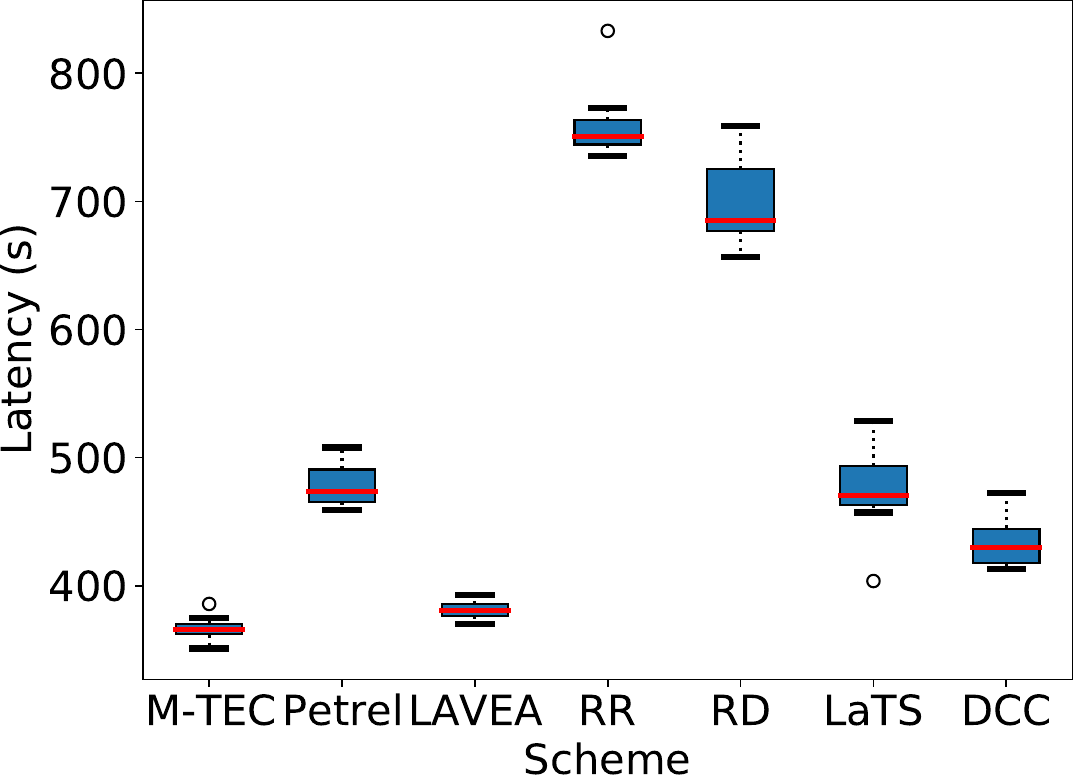}
\caption{End-to-end latency comparison of \name and baselines when probability of failure is not considered. \name outperforms all baselines.}
\label{fig:latency_5_deivce}
\vspace{-4 mm}
\end{figure}
To demonstrate that \name can lower the average end-to-end latency of an application instance, five edge devices are randomly selected from our testbed and added to our orchestration network. We programmed the initiator to send 100 application instance requests at random within 250s and performed the experiment ten times. By repeating the experiments, we are able to retrieve the average performance of the orchestration scheme as well as exam the overall system performance under randomly generated peak load, which was observed as more than 10 application instances within 5 seconds. Then, we compare the average end-to-end latency of each orchestration scheme's application instance.
Due to the fact that none of the baselines account for the probability of device failure or network variation, we did not induce any device failure in this experiment, and all five edge devices in the orchestration network are connected via LAN. As seen in Figure \ref{fig:latency_5_deivce}, \name outperforms all baselines. In addition, the performance of \name is stable as the average end-to-end latency does not vary significantly during 10 test cycles.

\subsubsection{\textbf{Latency vs. number of devices }} 
\begin{figure}[ht]
\includegraphics[width =0.95\columnwidth]{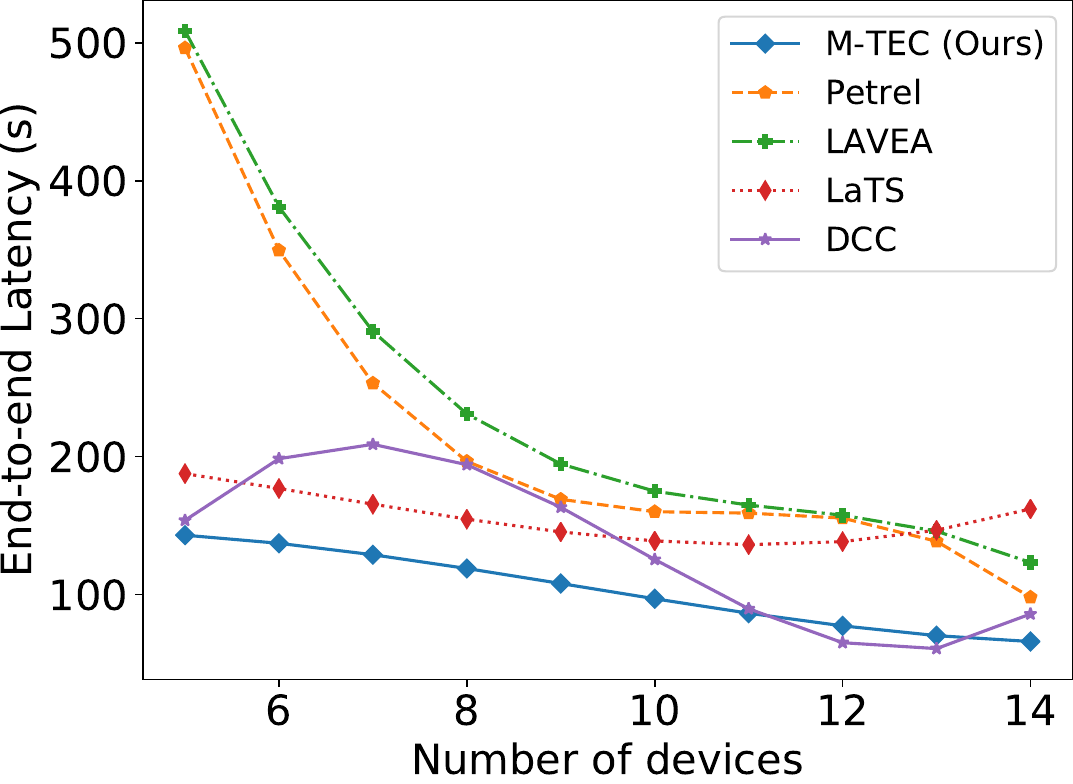}
\caption{Comparison of end-to-end latency between \name and baselines as the number of network edge devices increases. \name beats all baselines under varying device counts, particularly when there are few edge devices available on the network.}
\label{fig:latency_trend}
\vspace{-4 mm}
\end{figure}

~\\ \label{sec:lat_number}
As more devices are added to the orchestration network, the average end-to-end latency should ideally decrease. However, such an assumption is implausible and necessitates a vast number of devices in the orchestration network. Additionally, task distribution across several devices incurs communication latency. To demonstrate the benefits of \name, we conducted the latency test with varying numbers of edge devices in the network. Figure \ref{fig:latency_trend} demonstrates the trend of the average end-to-end latency of the mapreduce sort application for \name and four additional baselines. We eliminated the two intuitive baselines RR and RD from Figure \ref{fig:latency_5_deivce} since they are out of scale. The average end-to-end latency for \name and baselines reduces as the number of devices in the orchestration network increases. \name continues to exceed the second baseline by more than 50 percent when fourteen devices are added to the orchestration network. Moreover, we observe that DCC's performance fluctuates over the course of the test. This is because the device clustering method can generate clusters with uneven computation capabilities, making some of them the performance bottleneck. We would like to point out that as more devices are added to the network, the performance improvement starts decreasing and it introduces excessive orchestration overhead. Therefore, the number of devices in an initiator's network does not grow unbounded.

\subsection{\textbf{Probability of application failure}}
\begin{figure}[ht]
\includegraphics[width =0.95\columnwidth]{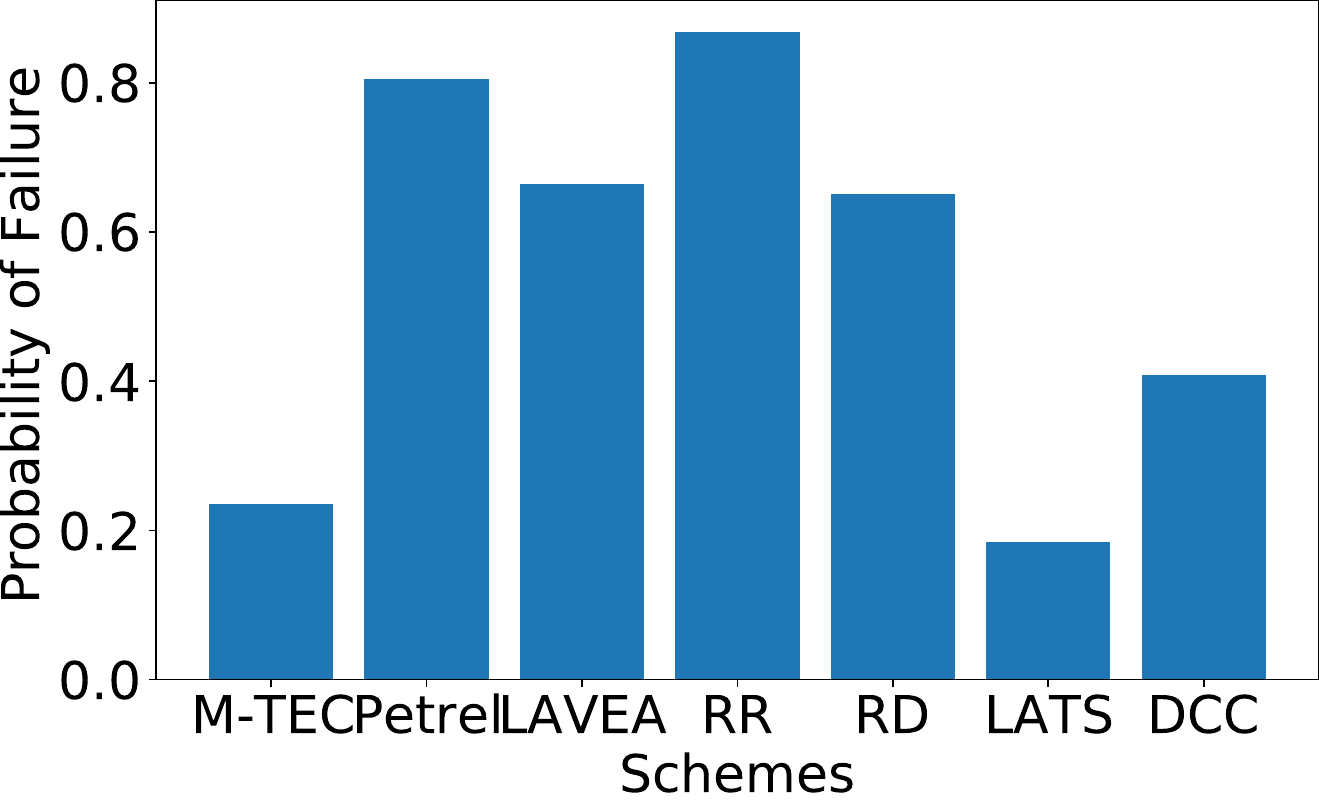}
\vspace{-3 mm}
\caption{Probability of failure comparison for application lightgbm, \name outperforms all baselines (except LaTS).}
\label{fig:lightgbm_pf}
\vspace{-3 mm}
\end{figure}

To examine the efficacy of \name in reducing the failure probability, we allowed each of the 14 testbed devices to fail according to an exponential distribution. Each device's failure rate $\lambda$ is carefully chosen, and it may be linked to real-world device failure data collected by us~\cite{mobility}. Figure~\ref{fig:lightgbm_pf} depicts the average probability of failure when executing 100 application instances in 250 seconds. We observe that \name reduces the probability of failure to approximately 20\%. We see that the performance of LaTS~\cite{lats} is equivalent to that of \name, however to achieve such a low probability of failure, LATS distributed the majority of work to devices with powerful computational capabilities. This allocation technique is not ideal since the failure of a single device can result in the failure of a significant number of applications.

\subsection{\textbf{Effect of dynamic network conditions}}
\begin{figure}[ht]
\centering
\includegraphics[width =0.95\columnwidth]{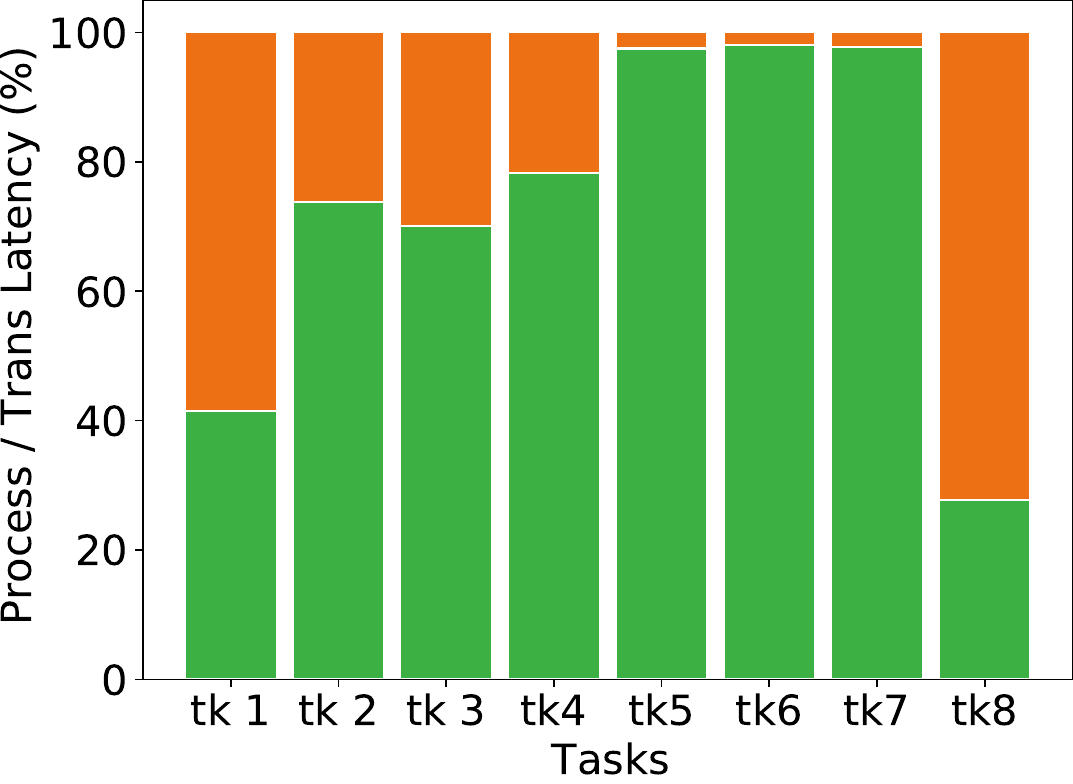}
\vspace{-3 mm}
\caption{Process (green) and transmission (orange) latency expressed as a proportion of total latency for a round robin distribution scheme in a CBRS 4G network. The transmission latency can comprise up to 60\% or more of the overall latency.}
\label{fig:process_trans}
\vspace{-1 mm}
\end{figure}

Transmission latency varies greatly in a heterogeneous computing platform where devices in the framework are connected to different networks in different locations. Figure \ref{fig:process_trans} depicts the process delay and transmission latency of the video analytics applications distributed in our testbed in a round-robin fashion. For certain tasks, such as splitting and transmitting the splitted video, transmission latency can account for up to 60\% of the overall latency.

\begin{figure}[ht]
\includegraphics[width =0.9\columnwidth]{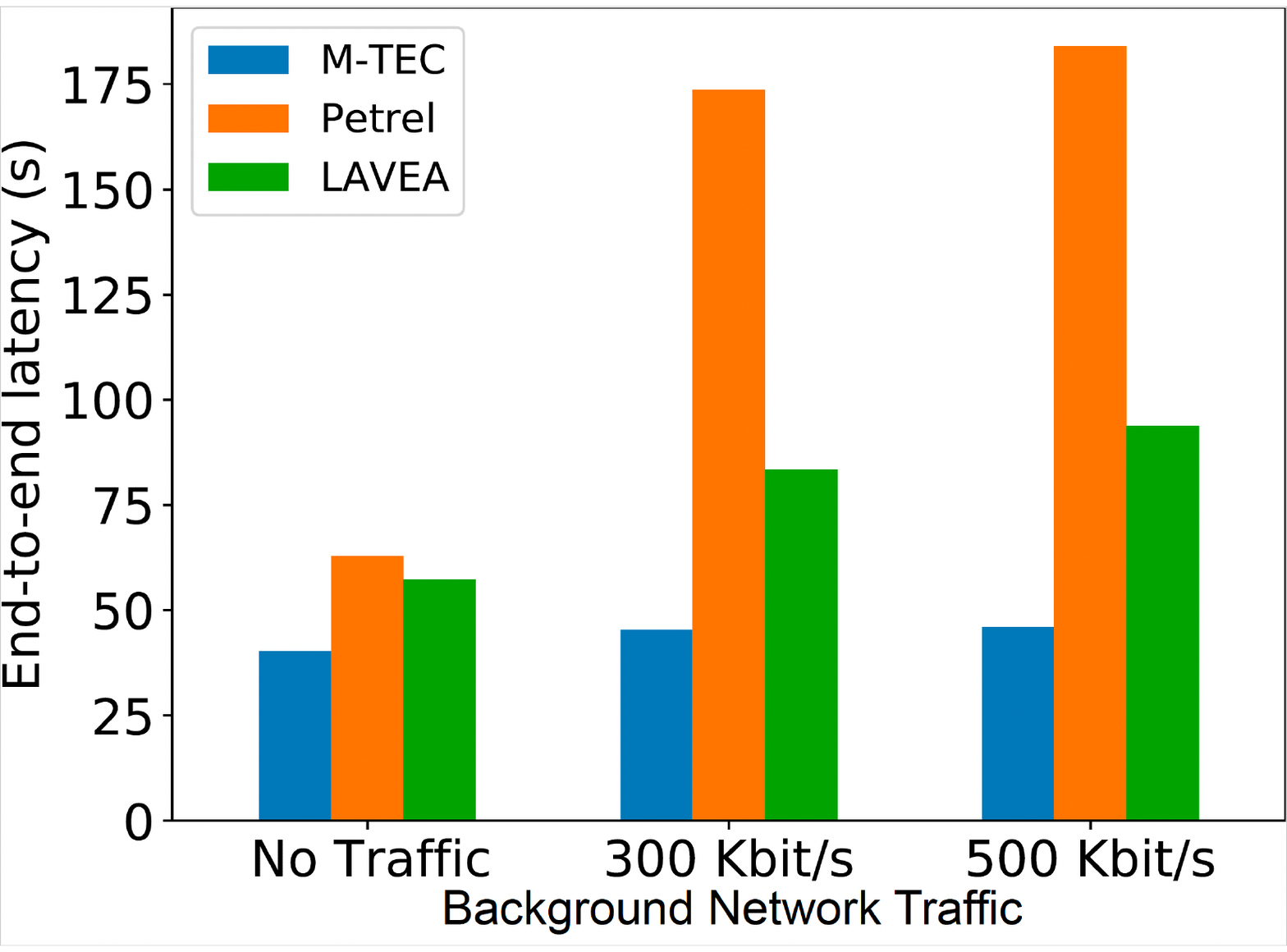}
\vspace{-5 mm}
\caption{End-to-end latency of video analytics application under 4G CBRS network with various network traffics}
\label{fig:network_traffic}
\end{figure}

To evaluate the performance of \name in a heterogeneous and dynamic network environment, we conduct a real test with the CBRS 4G network and inject network traffic into the wireless connections using \textit{iperf}. Figure \ref{fig:network_traffic} depicts the average end-to-end latency of a video analytics application under various network traffic scenarios. We compare \name to LAVEA~\cite{lavea} and Petrel~\cite{petrel}, the two top performers in the scenario with no traffic. When there is no traffic in the network, we can see that all three systems function similarly, with \name winning by a small margin. We notice, however, that \name is capable of adapting to the network condition and maintaining a rather consistent performance with less than 10\% deterioration when we begin to introduce traffic. However, neither Petrel nor LAVEA is able to do so, and their latency increases by more than 80\%.

\subsection{\textbf{Hyper-parameters $\alpha, \beta, \gamma$}}
\begin{figure}[t]
\centering
\includegraphics[width = .95\columnwidth]{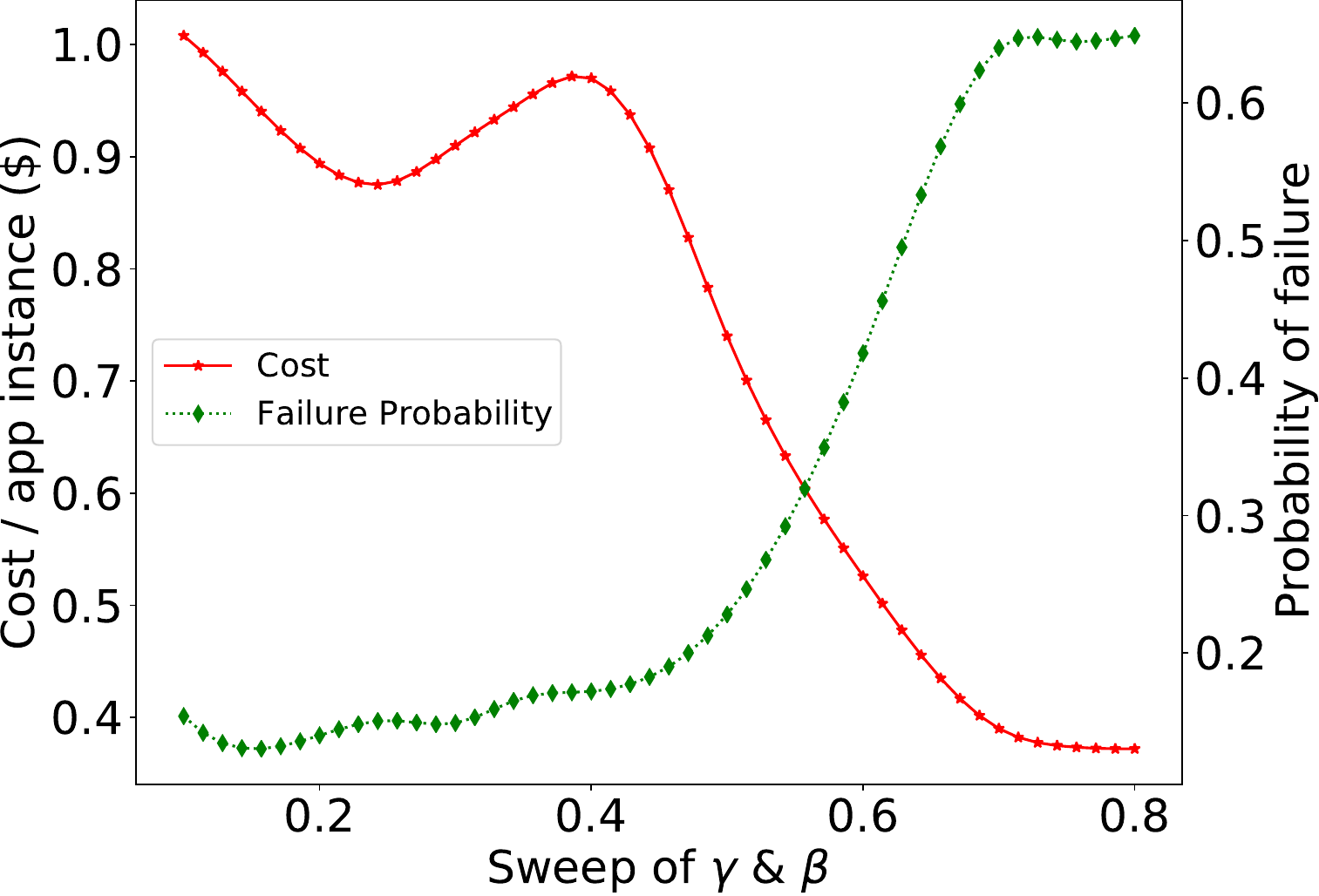}
\vspace{-3 mm}
\caption{Failure probability and cost sweep at $\alpha = 0.1$. On the x-axis, we have $\gamma$, which represents the weight assigned to the cost, and $\beta=1-\alpha-\gamma$. As more emphasis is placed on cost optimization, failure probability rises and the average cost per application instance falls.}
\label{fig:pf_cost}
\vspace{-3 mm}
\end{figure}

\begin{figure}[ht]
\centering
\includegraphics[width =.95\columnwidth]{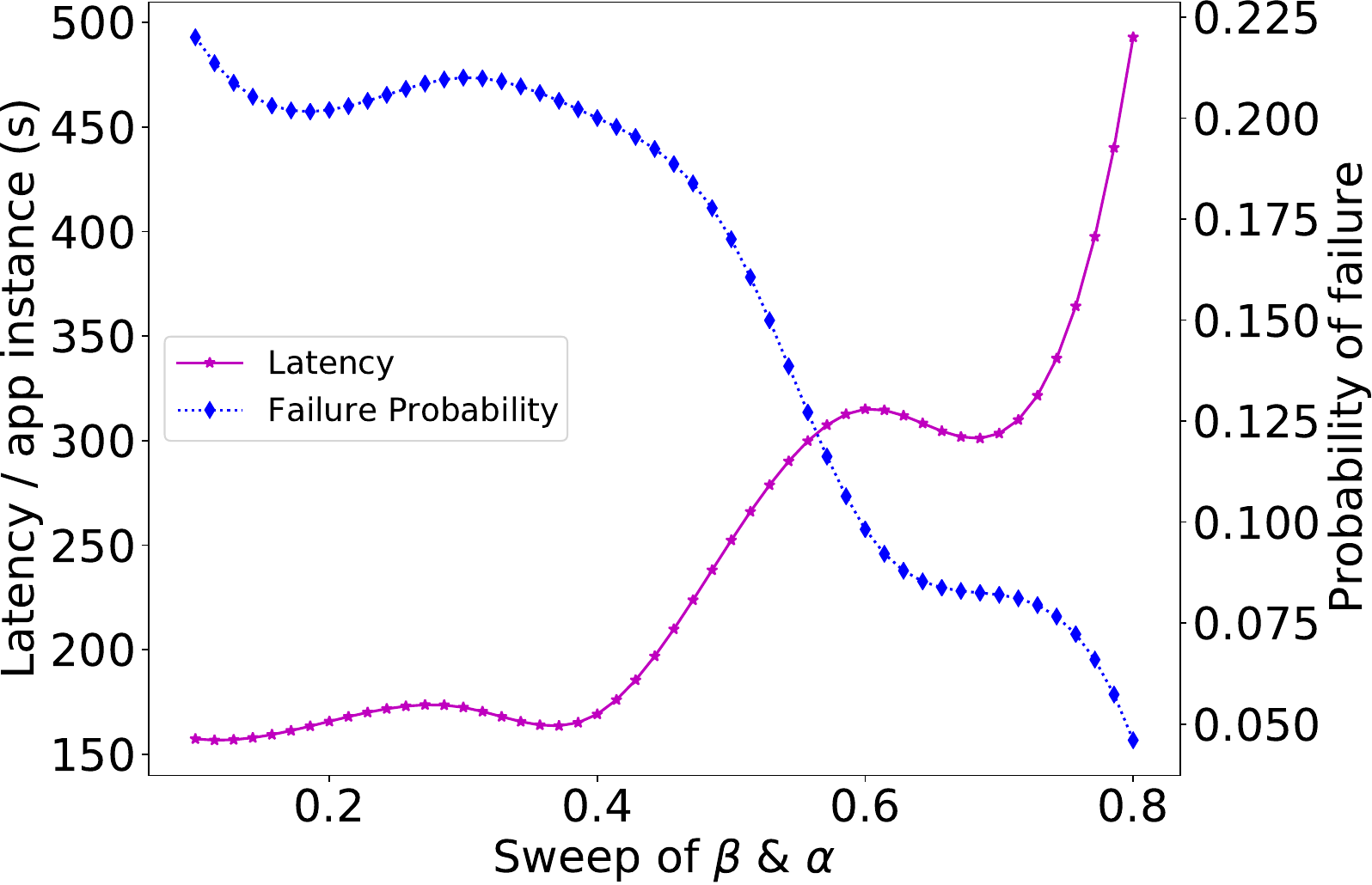}
\vspace{-3 mm}
\caption{Failure probability and latency sweep at $\gamma = 0.1$. On the x-axis is $\beta$, which corresponds to failure probability, and $\alpha=1-\beta-\gamma$. As more emphasis is placed on optimizing failure, the end-to-end latency of application rises and the failure probability falls.}
\label{fig:pf_latency}
\end{figure}

\begin{figure}[ht]
\centering
\includegraphics[width =.95\columnwidth]{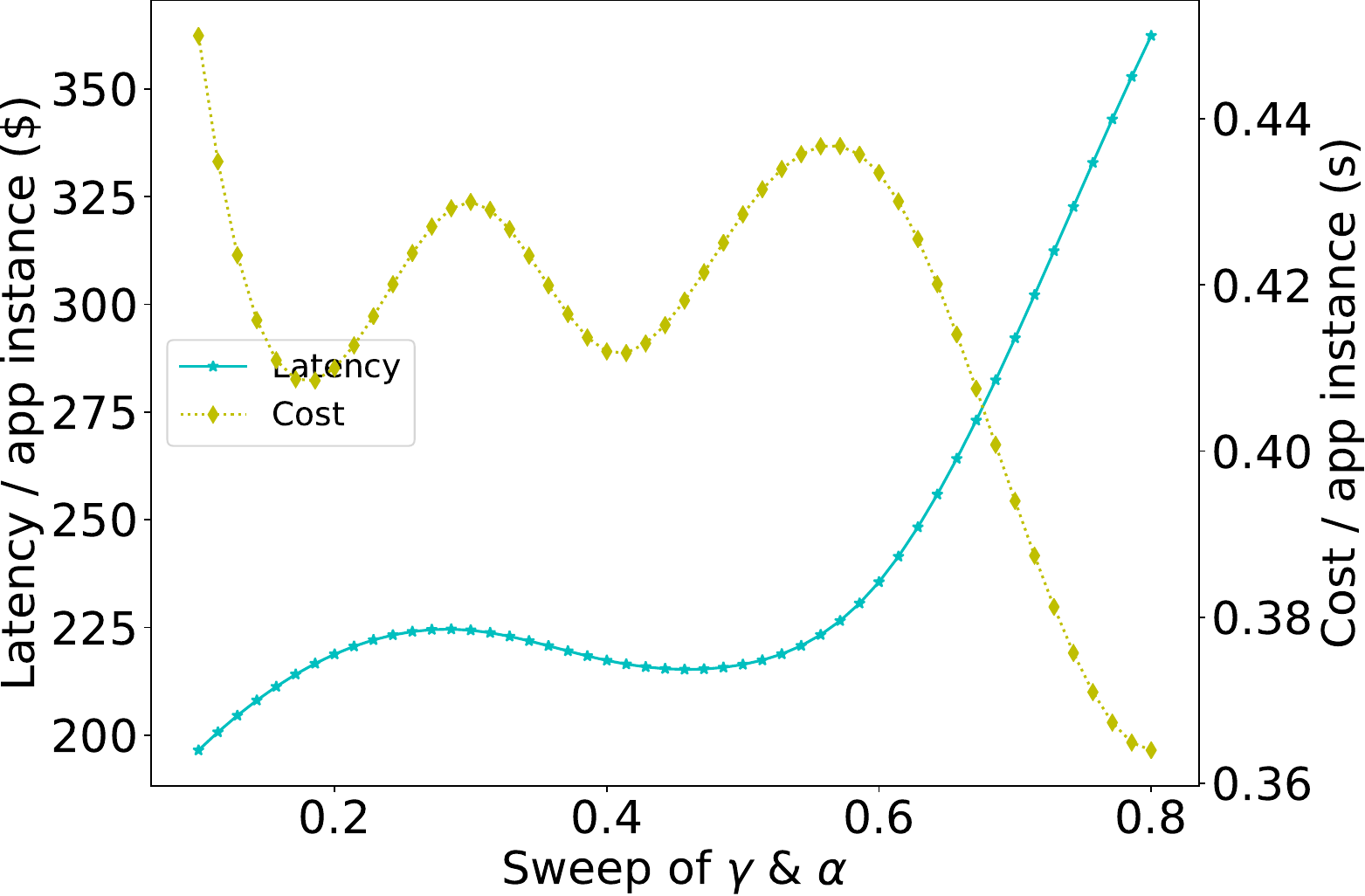}
\vspace{-3 mm}
\caption{Latency and cost sweep at $\beta = 0.1$. On the x-axis is $\gamma$, which corresponds to the weight assign to the cost, and $\alpha=1-\beta-\gamma$. As more emphasis is placed on cost optimization, the average end-to-end latency rises.}
\label{fig:cost_latency}
\vspace{-5 mm}
\end{figure}

We have now evaluated the performance of \name in terms of lowering latency and failure probability. Next, we demonstrate the performance of \name's joint optimization in \eqref{eqn:joint_optimization}. As depicted in Figures~\ref{fig:pf_cost}, \ref{fig:pf_latency}, and \ref{fig:cost_latency}, we did a sweep on the hyper parameters $\alpha$, $\beta$, and $\gamma$ to demonstrate the join optimization performance of \name.

In Figure \ref{fig:pf_cost}, we maintain a constant value of 0.1 for the hyper-parameter $\alpha$, which represents the weight associated with end-to-end latency. Then, a sweep was done on $\beta$ and $\gamma$. The result demonstrates that when the weight switches from failure probability ($\beta$) to cost ($\gamma$), \name gradually reduces the number of task replications, resulting in an increase in failure probability and a decrease in the average cost per application instance. At $\gamma=0.4$, instead of falling, the average cost per application momentarily increases. This is because, at this stage, there is a large decrease in end-to-end latency (associated with weight $\alpha=0.1$), which can result in greater overall performance than simply reducing the average cost. Consequently, we observe this transitory cost increase. The general pattern remains unchanged: as the probability of failure increases, the average cost drops.

Figure \ref{fig:pf_latency} depicts the sweep of hyper-parameters $\alpha$ and $\beta$, while $\gamma$ is held constant at 0.1. The result indicates that when $\alpha$ is given greater weight, the average end-to-end latency of application instances reduces and the probability of failure rises. Similarly, we also see that there is a temporary rise in average end-to-end latency (at $\alpha=0.3$, not as obvious as in Figure \ref{fig:pf_cost}), such behavior is caused by the cost weight $\gamma$. At around $\alpha=0.3$, the framework choose to sacrifice latency in return to reduce the cost, which gives an overall better performance.  


Figure \ref{fig:cost_latency}  illustrates the sweep of hyper-parameters $\alpha$ and $\gamma$, while holding $\beta$ constant at 0.1. Increasing the weight of $\alpha$ improves the average end-to-end latency of application instances while increasing the average cost per application. As observed, the latency plot varies during the experiment. This is due to the fact that when the cost is given less importance, devices with more computation capability and dependability are more likely to be chosen, resulting in a large reduction in failure probability. Consequently, the plot is marked by such variation.

\subsection{\textbf{Transmission speed error over-provision $\eta$}} \label{sec:eta_eva}

\begin{figure}[ht]
\centering
\includegraphics[width = .95\columnwidth]{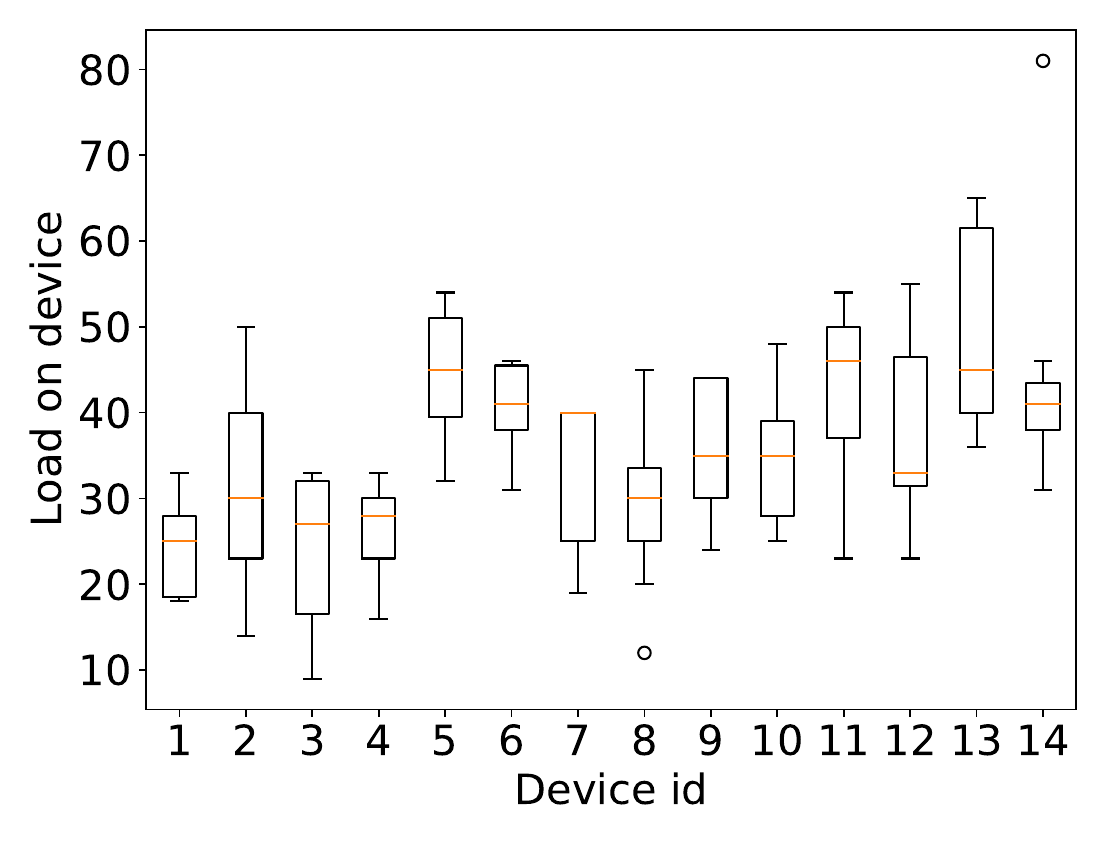}
\vspace{-7 mm}
\caption{The load on each devices under test when the transmission latency over-provision $\eta$ is set to varying values.}
\label{fig:eta_sweep}
\vspace{-3 mm}
\end{figure}

To determine how the transmission latency over-provision parameter $\eta$ influences task allocation, we gather the load on each device for $\eta$ values ranging from 1 to 3 with a 0.2-step increment for the matrix computation application. The outcome is depicted in Figure \ref{fig:eta_sweep}. At various $\eta$ values, there is a wide range of load on each device, but in the majority of cases, the load on a device falls between 30 and 60, which accounts for less than 15 percent of all tasks being dispatched (we are sending 100 application instances and each instance contains 5 tasks, so 500 tasks in total). As a result, changing the value of $\eta$ does not result in the aggregation of tasks on specific devices, which could lead to catastrophic failure.

%% file: Relatedwork.tex
\section{Related work} 
\label{sec:related_work}

Task scheduling in heterogeneous edge computing has been the subject of various studies in the literature. This section compares our work to those of our predecessors in this field.

\noindent\textbf{Multi-tier edge computing}: One of the primary goals of edge computing is to achieve minimal end-to-end latency in order to facilitate latency-sensitive applications. Several prior works~\cite{lats,lavea,petrel,mume, deep_decision} offered several scheduling algorithms for client-edge offloading that target various optimization objectives, such as latency, energy, and resource allocation, among others. We have demonstrated that \name outperforms LaTS~\cite{lats}, LAVEA~\cite{lavea}, and Petrel~\cite{petrel}, which are all latency-optimized schemes. There is a growing body of work~\cite{hetmec,hetmec2} on multi-layer edge computing. The majority of such multi-layer architectures adopt a bottom-up approach in which task offloading occurs only across layers or inside edge layers, but not within client layers. For instance, an end device can only offload tasks to edge servers or clouds, not to other end devices. Client-edge offloading is a subset of multi-layer offloading architectures with only two layers. Few works are evaluated within the layer task offloading~\cite{novel_work}. However, none of the literature addresses a framework that uses heterogeneous devices from diverse networks to create a dependable edge computing platform that minimizes end-to-end latency, failure probability, and cost.

\noindent\textbf{Dynamic and heterogeneous network}:
One of the bottlenecks for edge computing is network speed, which determines the rate at which tasks can be dispatched to their assigned edge nodes. Such transmission latency cannot be overlooked in a network environment that is dynamic. The vast majority of the available literature focuses solely on the network delay caused by dynamic network conditions~\cite{detour,mac}, sometimes caused by adversarial actions~\cite{mitra2019resilient}. \cite{lavea} briefly discussed the various performance of heterogeneous networks. In the course of our experiments, we have demonstrated that \name is able to adapt task allocation in diverse networks (Ethernet/CBRS 4G).

\noindent\textbf{Machine learning tasks on edge}:
Recent research attempts to offload machine learning and deep learning tasks to edge platforms~\cite{zhang2021nn,RTmDL}. \cite{RTmDL}, for instance, proposed RT-mDL, a framework that aims to reduce the rate of missed deadlines by employing a novel model scaling method and by utilizing joint model selection and task priority assignment to improve the performance and resource utilization of GPU/CPU for traffic light detection and sign recognition. This offloading approach is promising for some applications, but cannot be easily adapted to other applications, hence it lacks universality. In contrast, \name is capable of handling a variety of application types given the availability of their profiling information.

%% file: Discussion.tex
\section{Discussion and Future Work}
\label{sec:discussion}

In this section, we discuss the constraints of \name as well as various performance-enhancing options.

\noindent{First}, the current algorithm begins by comparing each incoming task to {\em all} available edge devices. When multiple edge devices are accessible, this procedure may result in a large orchestration overhead for simple tasks. Even though our research demonstrated that adding redundant edge devices does not improve performance (Section \ref{sec:lat_number}), we can still cluster edge devices based on their performance. The orchestration burden is thereby reduced from the quantity of devices to the quantity of clusters. For edge device clustering, any of various current approaches, such as~\cite{cluster1,cluster2,cluster3} or the Kmeans clustering used in DCC~\cite{novel_work} can be employed.

\noindent{Second}, the current offloading strategy delegated each task in the application as a whole to the edge devices. However, it is feasible that some tasks within an application may be particularly resource-intensive, therefore they may be subdivided into partial tasks and offloaded to other edge devices. There is an abundance of literature on partial offloading~\cite{partial1,partial2, partial3}.

\noindent{Third}, the current algorithm uses a hyper-parameter $\eta$ to estimate the network transmission error between the network speed measured at the time of task execution and the actual network speed at the time of task orchestration. This transmission error over-provision establishes a lower bound for the transmission rate and prevents task offloading that could result in long transmission latency under potentially adverse network conditions. This hyper-parameter can be a learnable parameter, meaning the algorithm can learn to set this hyper-parameter throughout the orchestration process depending on the past performance of the network estimation.



%% file: Conclusion.tex
\section{Conclusion} 
\label{sec:conclusion}

In this paper, we present \name, a novel multi-tier edge computing framework capable of running complex DAG-based applications. Importantly, \name incorporates client-to-client offloading in addition to the conventional client-edge and client-cloud offloading. We proposed an algorithm for the joint optimization of end-to-end latency, failure probability, and cost, taking into account the unpredictability of network and device failure. We evaluated \name on a real device testbed with a commercial CBRS 4G network using four applications spanning diverse DAG structures. We compared \name to four cutting-edge edge scheduling technologies, including DCC, LaTS, LAVEA, and Petrel. We find that \name reduces the end-to-end latency of applications by at least 8\%, the average probability of failure for applications by more than 40\%, and the average cost per application by around 12\% compared to the best baseline.